\documentclass[lettersize,journal]{IEEEtran}
\usepackage{amsmath,amsfonts}
\usepackage{algorithmic}
\usepackage{algorithm}
\usepackage{array}
\usepackage[caption=false,font=normalsize,labelfont=sf,textfont=sf]{subfig}
\usepackage{textcomp}
\usepackage{stfloats}
\usepackage{url}
\usepackage{verbatim}
\usepackage{graphicx}
\usepackage{cite}
\usepackage{breqn}
\usepackage{mathtools}

\usepackage{amsmath,amssymb,amsfonts}
\usepackage{algorithmic}
\usepackage{graphicx}
\usepackage{textcomp}
\usepackage{xcolor}

\usepackage[utf8]{inputenc} % allow utf-8 input
\usepackage[T1]{fontenc}    % use 8-bit T1 fonts
\usepackage{hyperref}       % hyperlinks
\usepackage{url}            % simple URL typesetting
\usepackage{booktabs}       % professional-quality tables
\usepackage{amsfonts}       % blackboard math symbols
\usepackage{nicefrac}       % compact symbols for 1/2, etc.
\usepackage{microtype}      % microtypography
\usepackage{amsmath,amsthm,amssymb}

\usepackage{svg,tikz}

\begin{document}

\title{Neural Decoding with Optimization of Node Activations}

\author{Eliya~Nachmani,
        Yair~Be'ery
        % <-this % stops a space
\thanks{E. Nachmani is with the School of Electrical Engineering, Tel-Aviv University, Tel-Aviv 6997801, Israel, and also with Facebook AI Research, Menlo Park, CA 94025, (e-mail: enk100@gmail.com)}
\thanks{Y. Be'ery is with the School of Electrical Engineering, Tel-Aviv University, Tel-Aviv 6997801, Israel (e-mail:ybeery@eng.tau.ac.il).}}

\maketitle

\begin{abstract}
The problem of maximum likelihood decoding with a neural decoder for error-correcting code is considered. It is shown that the neural decoder can be improved with two novel loss terms on the node's activations. The first loss term imposes a sparse constraint on the node's activations. Whereas, the second loss term tried to mimic the node's activations from a teacher decoder which has better performance. The proposed method has the same run time complexity and model size as the neural Belief Propagation decoder, while improving the decoding performance by up to $1.1dB$ on BCH codes.
\end{abstract}

\vspace{-0.5cm}
\section{Introduction}
Deep learning has become in last years an effective tool for communication tasks, for example, MIMO detection\cite{samuel2019learning, shlezinger2020deep}, modulation and demodulation \cite{ramjee2019fast, cohen2021learning, ramjee2020ensemble}, equalization\cite{caciularu2020unsupervised}, learning encoders and decoders \cite{weinberger2021generalization} designing new codes \cite{kim2018deepcode} which outperform codes for feedback channels \cite{ginzach2017random, ben2015gaussian}, feedback decoders \cite{shayovitz2021universal} and decoding\cite{nachmani2016learning, weinberger2021generalization}. 

Decoding error-correcting codes with novel deep learning techniques is an emerging research field. The Neural Belief Propagation (NBP) \cite{nachmani2016learning} was the first deep neural decoder that provide an improvement over the vanilla Belief Propagation decoder. Over the last few years, a large research effort to further improve the deep neural decoders was made to obtain better bit-error-rate decoding. Most of the previous work has tried to find better neural architecture to improve the decoder's performance. For example, Vasic et al.\cite{vasic2018learning} proposed a neural decoder that has activation functions that emulate message update functions. Their proposed method achieves higher throughput at the cost of additional decoding complexity. Nachmani et al. \cite{nachmani2019hyper} proposed an hypernetwork decoder that uses graph neural networks as a neural decoder that improves the NBP decoder with additional decoding complexity. Choukroun et al. \cite{choukroun2022error} suggested a Transformer neural decoder which improves the hypernetwork decoder at the cost of additional complexity. Chen et al. \cite{chen2021cyclically} showed that a neural decoder with a shift-invariant structure on the weights further improves the NBP results. Their method uses a bigger parity check matrix which increases the decoding complexity of the decoder. Another line of work tried to change the loss function to improve the performance. Lugosch et al. \cite{lugosch2018learning} introduced a novel syndrome loss function that penalizes the neural decoder for producing outputs that do not correspond to valid codewords. The proposed method yields decoders with improved bit-error-rate with no additional decoding complexity. A pruned version of the neural decoder has been proposed in \cite{buchberger2020pruning} which removes irrelevant check nodes. Another main research direction in the information theory community was to find sparse parity check matrices \cite{gallager1962low}, which resulted in better decoding performance. For example, in \cite{mackay1999good} very sparse matrices achieve performance that is close to the Shannon limit. The sparse parity check matrices result in a sparse Tanner graph which indicates that sparse constraint may improve the neural decoding architectures.

As we can observe, previous work have tried to improve the results by changing the neural architecture, the loss function at the output layer or finding better sparse parity check matrix. Therefore, one can ask whether a direct node optimization can further improve the results, since it gives direct gradient updates to the neural decoder that produce better learning. In this paper we propose to use two novel loss terms that optimize the decoders nodes: (i) sparse node loss and (ii) knowledge distillation loss. Sparse activation regularization \cite{kurtz2020inducing} is a well known technique that imposes sparse constrained on the activation with new loss function term. Since we know that sparse parity check matrices lead to better decoding performance, we propose teaching the neural network decoder with sparse nodes. Therefore, the proposed neural sparse decoder behaves as a neural decoder with sparse parity check matrix. As expected, our simulation shows that the learned neural sparse decoders gives large improvement over the baseline. Knowledge distillation is a basic technique in deep learning where one uses a teacher network to guide the training of a smaller student neural network \cite{hinton2015distilling}. We propose to use the knowledge distillation method by an expert teacher network to constraint the nodes of the proposed neural decoder. The student tries to decode the transmitted codeword with a novel loss term that mimics the teacher node activations. Our simulations shows that we can further gain improvement when using the knowledge distillation loss.

Since our method adds new loss terms to the training, the method preserves the same complexity run-time as the original neural decoder. Furthermore, our method does not have additional weights and has the same model size as the baseline neural decoder. Our paper makes the following contributions:
\begin{itemize}
    \item We present novel sparse node activation loss. 
    \item We present a knowledge distillation loss term.
    \item We demonstrate that each of the new loss terms (and both combined) improves the results of the baseline methods by a large margin without adding computational complexity.
\end{itemize}

\vspace{-0.25cm}
\section{Background}
In this section, we will briefly explain the Belief Propagation (BP) decoder for a communication scheme for Binary Phase Shift Keying (BPSK) and binary code. We will use similar notations as in Lugosch et al. \cite{lugosch2017neural} and recap the offset min-sum BP decoder \cite{chen2005reduced} as well as the neural min-sum decoder \cite{lugosch2017neural}.  
Denote the block code length by $n$, the transmitted and received information vector as $X \in \mathbb{R}^n$ and $Y \in \mathbb{R}^n$ respectively. The transmitted vector \textcolor{black}{$X$} has $-1,+1$ values since we consider BPSK modulation. The received vector $Y$ at the receiver is given by $Y=X+Z$, where $Z \in \mathbb{R}^n$ is the channel noise. 
\textcolor{black}{For additive white Gaussian noise (AWGN) channel, the input vector to the BP decoder is the log-likelihood-ratio (LLR). The $v$'th element $l_v$ of the LLR vector is given by:
\begin{equation}
    l_v = \log\frac{\Pr\left(Y_v = y_v| X_v=-1\right)}{\Pr\left(Y_v = y_v| X_v=+1\right)} = \frac{2y_v}{\sigma^2}
\end{equation}  
where $y_v$ is the channel output, $\sigma^2$ is the known channel noise variance, and $X_v$ is the $v$'th element of the transmitted vector $X$.}
The BP decoder exchanged messages between variable and check nodes layers iteratively. The variable and check layers obtain from the Tanner graph of the parity check matrix. The variable and check layer contain variable and check nodes respectively. An edge $e$ connects $c$ check nodes to $v$ variable node if the \textcolor{black}{$(c,v)$} entry in the parity check matrix contains $1$.
The message from the variable node $v$ to check node $c$ is given by:
\begin{equation}
    \mu_{v,c}(t) = l_v + \sum_{c' \in \mathcal{N}(v) \backslash c} \mu_{c',v}(t-1)
\end{equation}
where $t$ is the number of iterations, $\mathcal{N}(v)$ is the neighbors' check nodes of the variable node $v$. For $t=0$, the initial message is zero, i.e. $\mu_{v,c}(0) = 0 $.
The message from check node $c$ to the variable node $v$ is given by:
\begin{multline}\label{eq:eq6}
         \mu_{c,v}(t) = \max\left( \min_{v' \in \mathcal{M}(c) \backslash v }  (|\mu_{v',c}(t)|) - \beta, 0 \right) \cdot \\ \cdot \prod_{v' \in \mathcal{M}(c) \backslash v } \textrm{sign} (\mu_{v',c}(t))
\end{multline}
where $\mathcal{M}(c)$ is the neighbors' variable nodes of the check node $c$ and $\beta$ are the correlation offset. Note that the message in Eq.\ref{eq:eq6} is an approximation of the optimal message, with reduced complexity.
Lugosch et al. \cite{lugosch2017neural} proposed to replace Eq.\ref{eq:eq6} of the offset min-sum BP decoder with:
\begin{multline}\label{eq:eq14}
         \mu_{c,v}(t) = \max\left( \min_{v' \in \mathcal{M}(c) \backslash v }  (|\mu_{v',c}(t)|) - \beta_{c,v}(t), 0 \right) \cdot \\ \cdot \prod_{v' \in \mathcal{M}(c) \backslash v } \textrm{sign} (\mu_{v',c}(t))
\end{multline}
where $\beta_{c,v}(t)$ are learnable parameters for each edge from check node $c$ to variable node $v$.
After $t$ iterations, the soft output vector $s_v(t)$ is given by:
\begin{equation}
        s_v(t) = l_v + \sum_{c' \in \mathcal{N}(v)}  \mu_{c',v}(t)
\end{equation}

The output bits at the final iteration $T$ can be extracted with a hard slicer:
\begin{equation}
        \hat{x}_v(t) = 
\begin{cases}
    1,& \text{if } s_v(t) > 0
    \\0,              & \text{otherwise.}
\end{cases}
\end{equation}

The neural min-sum decoder uses a stochastic gradient descent in order to find $\beta_{c,v}(t)$ parameters that improve the performance of the offset min-sum BP decoder. The loss function is cross-entropy:
\begin{equation}
    \label{eq:ce}
    \mathcal{L}_{ce}=-\frac{1}{n}\sum_v B_v \textrm{log} (\sigma(s_v)) + (1-B_v) \textrm{log} (1 - \sigma(s_v))
\end{equation}
where $\sigma$ is the sigmoid activation function, and $B_v=(1-X_v)/2$ is the bit corresponding to transmitted symbol $X_v$.

\vspace{-0.25cm}
\section{Method}
We suggest adding sparse node activation loss and knowledge distillation loss to the training procedure of the neural min-sum decoder \cite{lugosch2017neural}. The sparse node activation loss is obtained by adding an $L_p$ norm on the check and variable nodes. Furthermore, the knowledge distillation loss imposes an additional constraint on the check nodes. We suggest using the knowledge distillation method by incorporating a teacher decoder into the training procedure. We recommend using the following component: (i) A teacher decoder and (ii) A student neural decoder. At test time, decoding is performed using the student neural decoder.
\begin{figure*}
    \vspace{-0.6cm}
    \centering
    \includegraphics[width=0.8\textwidth]{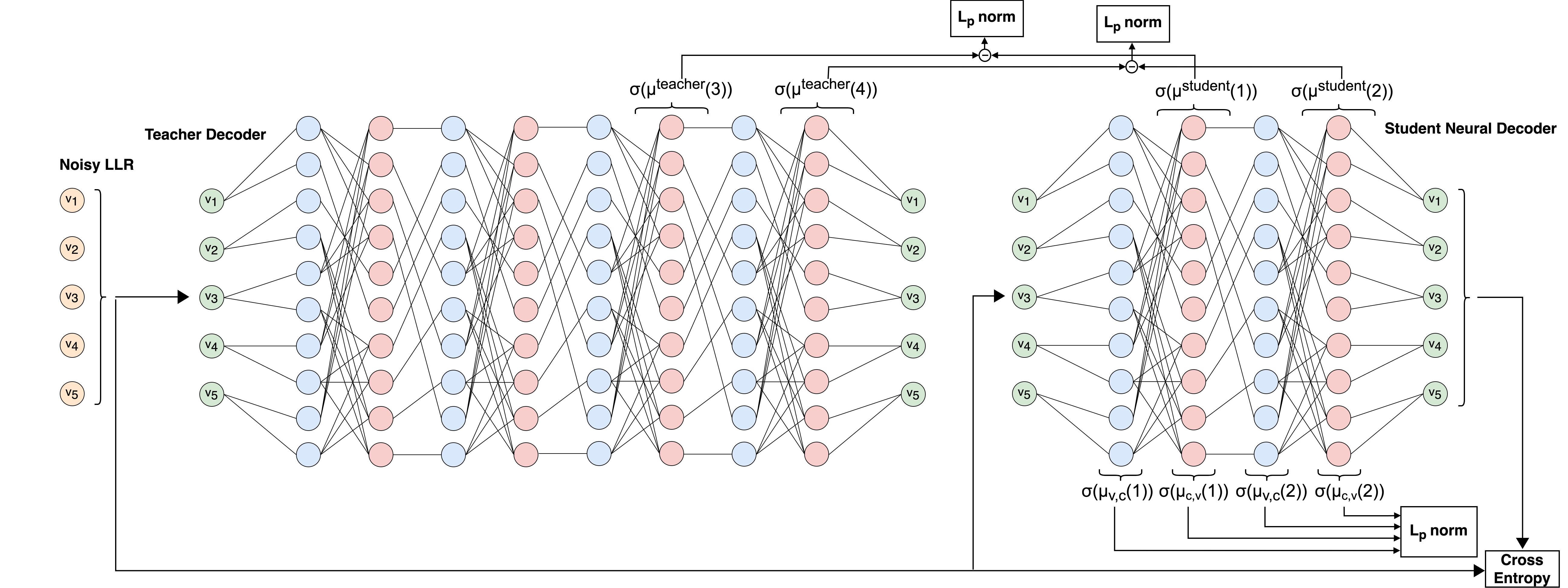}
    \vspace{-0.2cm}
    \caption{Our method. The Tanner graph for a linear block code with $n=5$. The teacher decoder has $T_{teacher}=4$ iterations whereas the student neural decoder has $T_{student}=2$ iterations. The look ahead parameter was set to $t_o=2$.}
    \label{fig:decoder}
    \vspace{-0.6cm}
\end{figure*}
\vspace{-0.4cm}
\subsection{The Architecture}
Our proposed method uses a min-sum decoder as a teacher (i.e. without any training), and a neural min-sum decoder as a student decoder. The specific implementation of the student neural network is based on the neural min-sum decoder \cite{lugosch2017neural}. It should be noted that one can also use different Neural Belief propagation decoders like the neural Sum-Product \cite{nachmani2016learning} as the neural decoder. A diagram of the proposed method is shown in Fig.~\ref{fig:decoder}. Denote by $T_{teacher}$ and $T_{student}$ the number of iterations in the teacher and student decoders respectively. Since the teacher decoder should have better bit-error-rate results than the student neural decoder, we set $T_{teacher} > T_{student}$. 
\vspace{-0.4cm}
\subsection{The Knowledge Distillation Loss Term}
\textcolor{black}{We propose to constrain the nodes of the neural decoder using the knowledge distillation method. The knowledge distillation method uses an expert teacher network (vanilla min-sum decoder, i.e. without learnable parameters) and a novel loss term that imitates the teacher's node activations. Since the teacher network has more layers, imitating the teacher's node activations, will result in lower decoding error. Therefore,} we propose a new loss function to guide the training of the neural decoders. Denote by $\mu_{c,v}^{teacher}(t)$ and $\mu_{c,v}^{student}(t)$ the teacher and student check node messages at iteration $t$, respectively.  
The knowledge distillation training tried to transfer knowledge from the teacher network into the student network. In other words, the student network messages $\mu_{c,v}^{student}(t)$ at iteration $t$ tried to mimic the teacher network messages $\mu_{c,v}^{teacher}(t+t_o)$ at iteration $t+t_o$. Where $t_o$ is a look ahead parameter, and it controls the expert level of the teacher network.
At training, the knowledge distillation method will generate a vector of LLR values with the same encoded bits vector for the teacher and the student networks, and apply the following loss function:
\begin{equation}
    \mathcal{L}_{kd}(t) = \sum_{v=1}^n \sum_{c'\in \mathcal{N}(v)} \left \| \sigma(\mu_{c',v}^{teacher}(t+t_o)) -  \sigma(\mu_{c',v}^{student}(t)) \right \|_p ^p
\end{equation}
where $p$ is the norm order and $\sigma$ is the sigmoid activation function. We apply the sigmoid function $\sigma$ in order to bound the messages $\mu_{c,v}^{student}(t)$ to range $[0,1]$.
The overall distillation loss function $\mathcal{L}_{kd}$ is given by the summing over \textcolor{black}{$T_{student}$} iterations:
\begin{equation}
    \label{eq:kd}
    \mathcal{L}_{kd} = \sum_{t=1}^{T_{student}} \mathcal{L}_d(t)
\end{equation}
\vspace{-0.5cm}
\subsection{The Sparse Node Activation Loss Term} 
\textcolor{black}{Sparse parity check matrix leads to higher decoding performance \cite{mackay1999good}. Therefore, we suggest a neural network decoder with sparse node activations. The suggested neural sparse decoder functions as a neural decoder with a sparse parity check matrix. Sparse node activations can be achieved with a sparse loss term.} The proposed sparse loss term is obtained by using a $L_p$ norm on variable and check nodes over the student network:
\begin{multline}
    \mathcal{L}_s(t) = \sum_{v=1}^n \sum_{c'\in \mathcal{N}(v)} \left \|\sigma(\mu_{c',v}^{student}(t)) \right \|_p ^p + \\ + \sum_{c=1}^{n-k} \sum_{v'\in \mathcal{M}(c)} \left \|\sigma(\mu_{v',c}^{student}(t)) \right \|_p^p
\end{multline}
where we use the same $p$ value as in the knowledge distillation loss terms. {We observe that the $p$ parameter is important for successful training since larger $p$ increases the gradient. We ran experiments to check the best $p$ value for training; further explanation will be given in the results section}. Furthermore, we apply the sigmoid $\sigma$ function over the variables and check nodes in order to squeeze the range to $[0,1]$.
Summing the sparse loss term at each iteration $t$, we obtain 
the overall sparse loss term:
\begin{equation}
    \label{eq:sa}
    \mathcal{L}_s = \sum_{t=1}^{T_{student}} \mathcal{L}_s(t)
\end{equation}
\vspace{-0.5cm}
\subsection{The Combined Loss Term}
The overall loss is a linear combination of the cross entropy loss Eq.\ref{eq:ce}, the knowledge distillation loss Eq.\ref{eq:kd} and the sparse node activation loss Eq.\ref{eq:sa} $\mathcal{L} = \mathcal{L}_{ce} + \alpha \mathcal{L}_{kd}+ \gamma \mathcal{L}_s$ where we set $\alpha=1$ and $\gamma=0.01$.
\vspace{-0.5cm}
\textcolor{black}{
\subsection{Complexity Analysis} 
The inference complexity is similar to the neural min-sum decoder. Both have the same number of layer and the same number of non-zero weights in the Tanner graph.
}
\begin{figure*}[t]
\vspace{-0.8cm}
\centering
\begin{tabular}{c@{~}c}
\includegraphics[width=.375\textwidth]{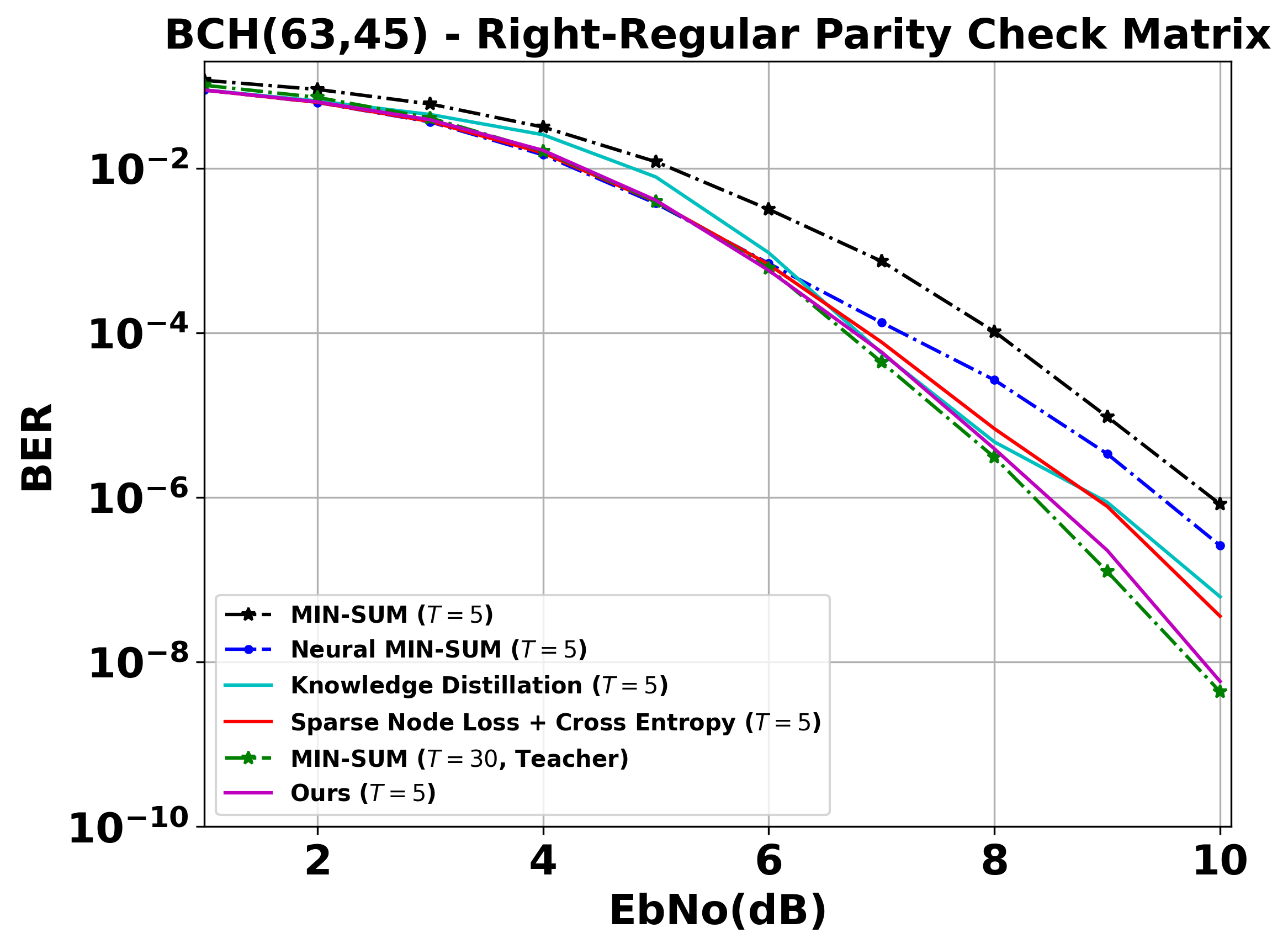}& 
\includegraphics[width=.375\textwidth]{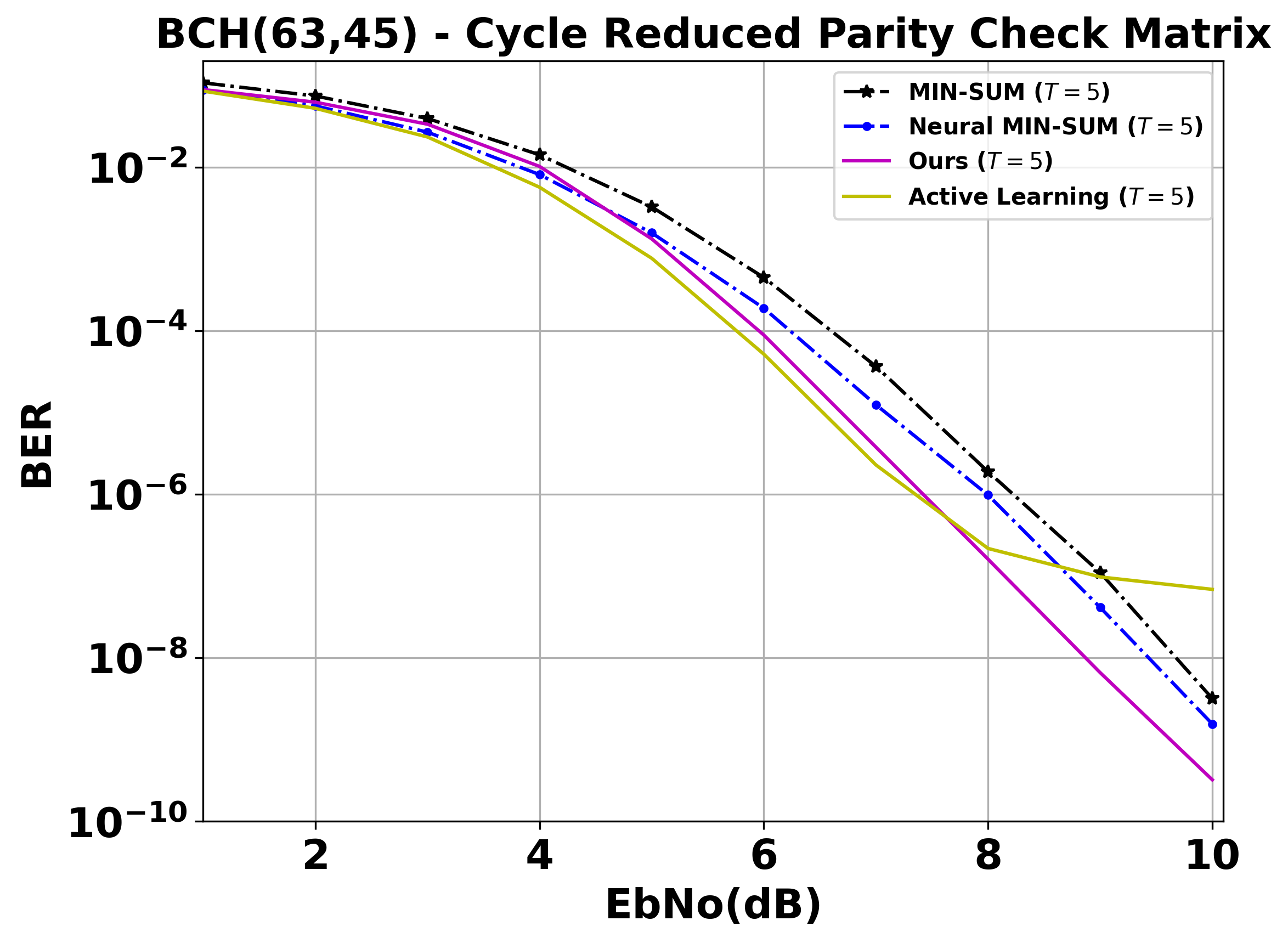}\\
(a) & (b)\\
\includegraphics[width=.36\textwidth]{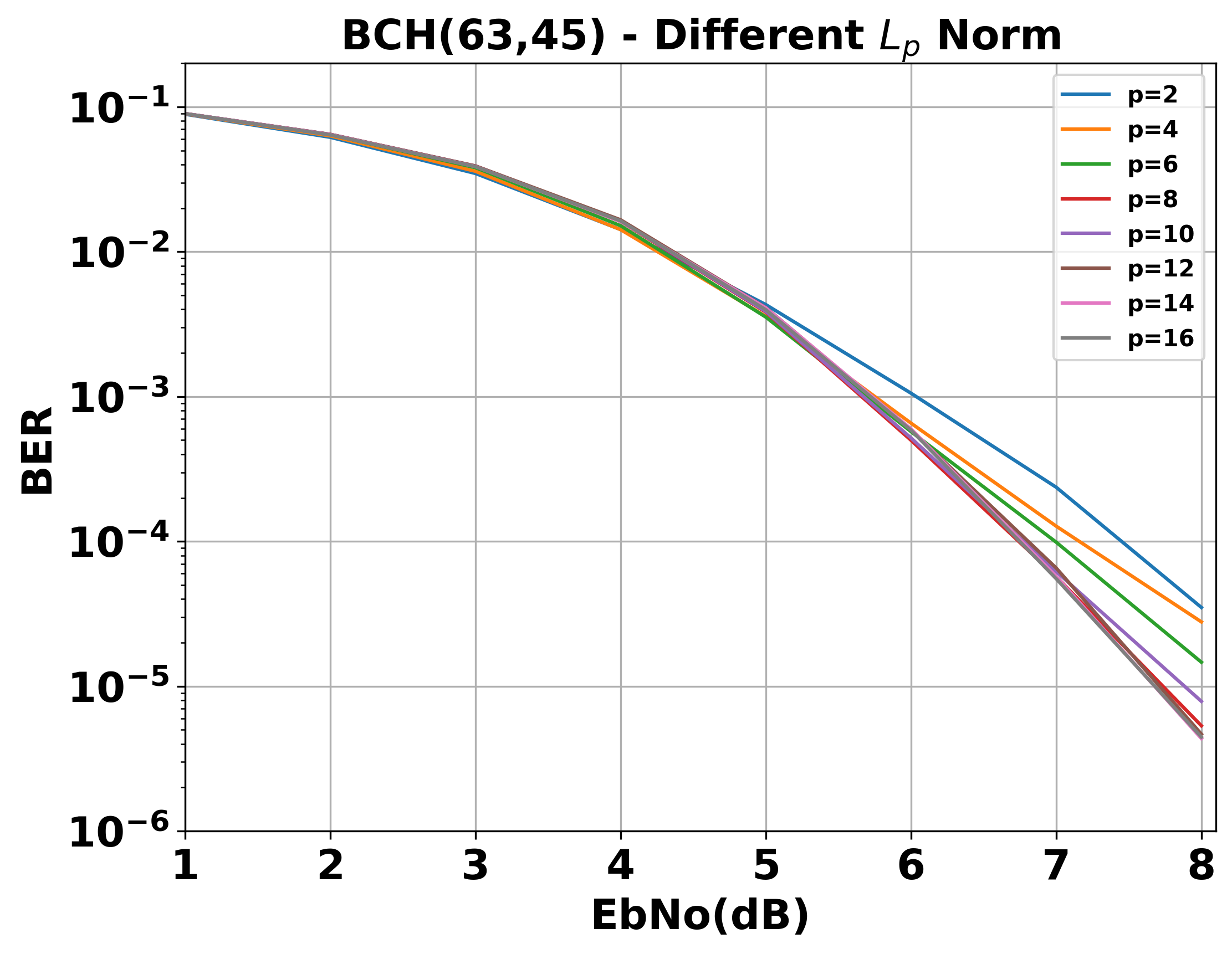}& 
\includegraphics[width=.36\textwidth]{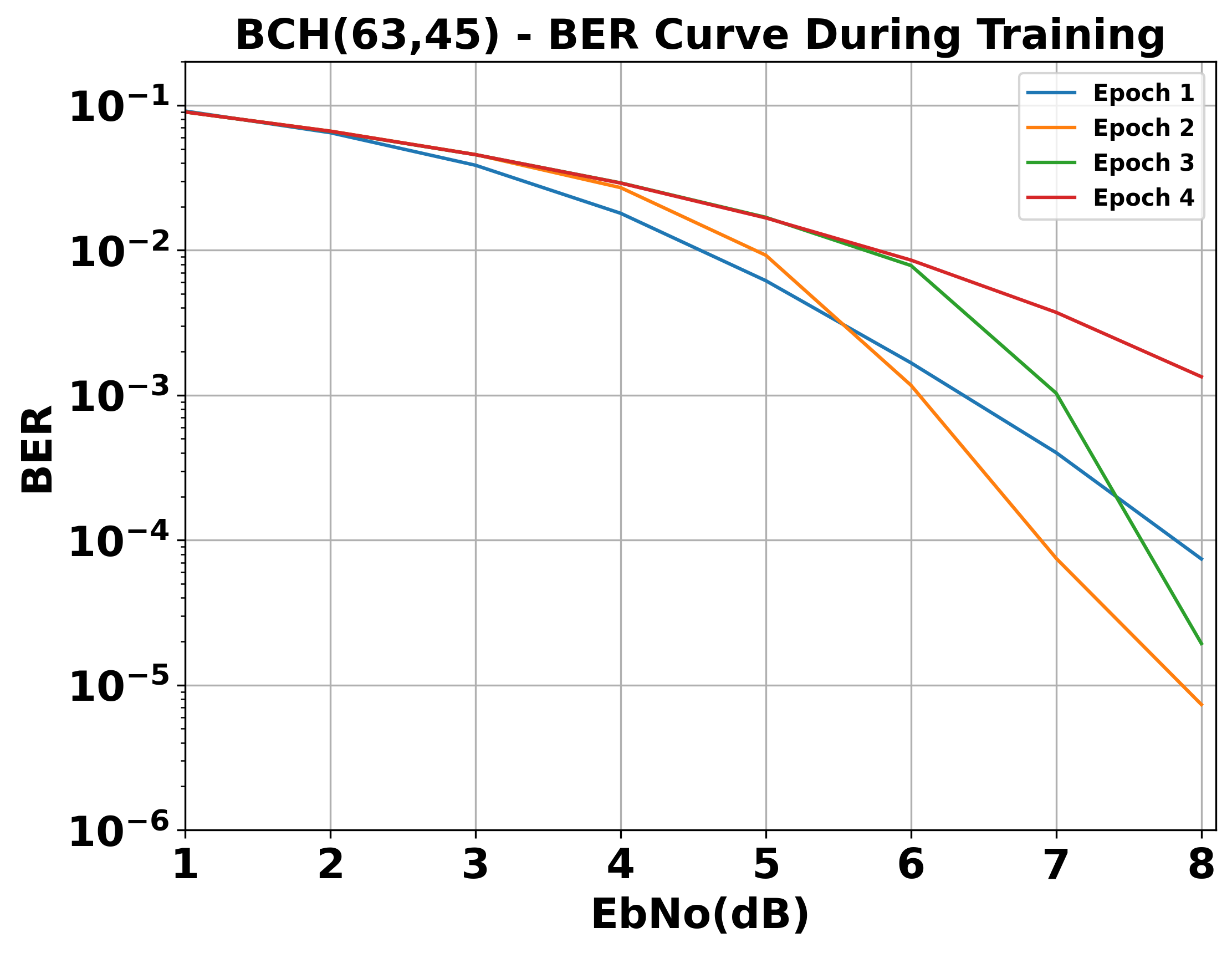}\\
(c) & (d)\\
\end{tabular}
\caption{BER for various values of SNR for BCH codes. (a) BCH (63,45) with a right-regular parity check matrix, (b) BCH (63,45) trained with a cycle reduced parity check matrix, (c) BCH (63,45) with a right-regular parity check matrix trained with different $p$ values. (d) BCH (63,45) with a right-regular parity check matrix, BER curve during training sparse node activation loss.}
\label{fig:ber_snr}
\vspace{-0.4cm}
\end{figure*}

\begin{figure}[t]
\centering
\includegraphics[width=.36\textwidth]{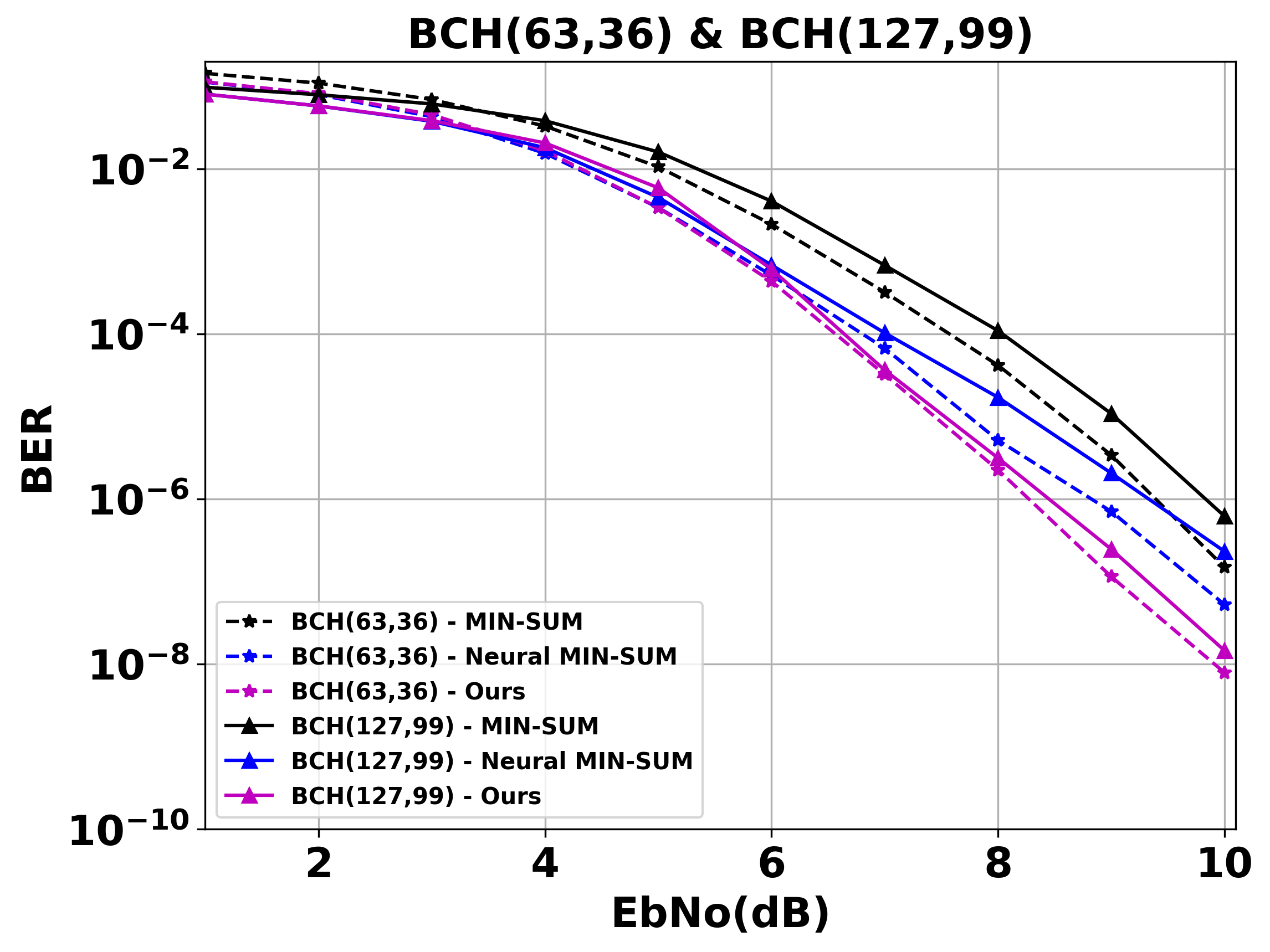}
\caption{BER for various values of SNR for BCH (63,36) \& BCH (127,99) codes trained with a cycle reduced parity check matrix. The number of iterations set to $T=5$ in all decoders.}
\label{fig:ber_snr_both}
\vspace{-0.6cm}
\end{figure}

\vspace{-0.25cm}
\section{Results}
The proposed architecture trained on Bose–Chaudhuri–Hocquenghem (BCH) codes \cite{bose1960class}. The parity check matrices and the generator matrices are taken from \cite{channelcodes} except for the cycle reduced parity check matrices that are based on \cite{cycle_reduce}. For the training and test set, we randomly generated codewords that transmitted over an additive white Gaussian noise (AWGN) channel. Each batch contained $360$ examples. Furthermore, the batch contained codewords with SNR values of $1dB,...,8dB$ with $45$ examples per SNR value. Moreover, the loss function is the mean value of the combined loss $\mathcal{L}$ for each example in the batch. The teacher network had $T_{teacher}=30$ layers while the student network trained with $T_{student}=5$ iterations, the look-ahead parameter is $t_o=25$. The norm order $p$ was set to $12$. The learning rate was set to $0.01$. We ran the Monte-Carlo simulation and obtained the Bit-Error-Rate (BER) for SNR values from $1dB$ to $10dB$. 

In Fig.\ref{fig:ber_snr}(a) we provide the BER figures for BCH (63,45) code with a right-regular parity check matrix. As can be seen, our proposed decoder improves the Min-Sum Belief Propagation decoder by up to $1.5dB$ for large SNR values ($6-10dB$), and $1dB$ for medium and small SNR values ($1-5dB$). Moreover, our proposed decoder improves the Neural Belief Propagation (NBP) \cite{lugosch2017neural} decoder by $1.1dB$ for larger SNR values ($8-10dB$), and up to $0.75dB$ improvement for medium SNR values ($6-7dB$). For smaller SNR values ($1-5dB$) our proposed method achieves the same decoding results as the Neural Belief Propagation (NBP) decoder. It should be noted that our method has the same run-time complexity and model size as the Neural Belief Propagation (NBP) decoder. 
Fig.\ref{fig:ber_snr}(b) provides the BER figures for BCH (63,45) code with a cycle reduced parity check matrix. As can be seen, our proposed method improves the Min-Sum Belief Propagation decoder by up to $0.75dB$ for large SNR values. Moreover, the proposed method improved the Neural Belief Propagation (NBP) \cite{lugosch2017neural} decoder by $0.6dB$ for larger SNR values. We also compared our method to the Deep Active Learning Decoder \cite{be2019active}, and found that it improves by the large gap $1.8dB$ the Deep Active Learning Decoder. Moreover, the Deep Active Learning Decoder is based on the Sum-Product decoder which has better results than the Min-Sum decoder which is the underline decoder for our method. For the low SNR regime, the Deep Active Learning Decoder achieves better BER with an improvement of up to $0.3dB$. However, it comes with a cost of error-floor at the high SNR regime. It is left for future research to combine the active learning decoder with our method for better BER in the low SNR regime.
In Fig.\ref{fig:ber_snr_both} we provide the BER for BCH (63,36) code with a cycle reduced parity check matrix. We can observe a large improvement of up to $1.1dB$ for large SNR values when compared to the Min-Sum Belief Propagation decoder. When compared to the Neural Belief Propagation (NBP) \cite{lugosch2017neural} decoder, our proposed method improves the results by $0.7dB$ for larger SNR values. For small SNR values, our proposed method has a small degradation ($0.2dB$) in the performance when compared to the Neural Belief Propagation (NBP) decoder. The source of the degradation is the knowledge distillation loss term which gets worse results in the low SNR regime (see the ablation section for more details). Possible sources for the degradation can be explained by the fact that the knowledge distillation loss term does not use the ground truth bits at training, while the Neural Belief Propagation (NBP) decoder uses this additional information.
Fig.\ref{fig:ber_snr_both} presents the BER for larger BCH code (127,99) with a cycle reduced parity check matrix. As can be observed from the figure, our method improves the Min-Sum Belief Propagation decoder by $1.4dB$ for high SNR values. Furthermore, it improves the performance of the Neural Belief Propagation (NBP) by up to $1.0dB$ in the high SNR regime. 

\noindent{\bf Ablation Analysis\quad} In Fig.\ref{fig:ber_snr}(a) we evaluate the contribution of different components of our method with an ablation analysis for the BCH(63,45) code. We compared our method with (i) a neural decoder that trains only with the knowledge distillation loss (cyan curve), (ii) a neural decoder that trains with the sparse node activation loss (red curve), and (iii) the teacher min-sum decoder with $T=30$ iterations (green curve). As we can observe, our method (magenta curve) achieves the best results which demonstrates the advantage of our complete method. Furthermore, we can observe that without the sparse node activation loss (i) and without the knowledge distillation loss (ii) we have a degradation of $0.5dB$ and $0.3dB$ respectively. This shows that the sparse node activation loss gives more improvement than that knowledge distillation loss. The teacher min-sum decoder (iii) can be regarded as lower bound to (i). We can observe that training only with the knowledge distillation loss (i) leads to degradation in the low SNR regime ($3-6dB$). The degradation is alleviated when training with the cross-entropy loss.

\noindent{\bf Training With Different $L_p$ Norms\quad} Fig.\ref{fig:ber_snr}(c) provides the BER curve for different $L_p$ norms. We use the BCH (63,45) code with a right-regular parity check matrix, and test it for $p=2,4,6,8,10,12,14,16$. As we can observe the curve congregates for $p$ value that is greater than $12$. Therefore, we choose to use $p=12$ in our simulations. \textcolor{black}{Furthermore, we observe that $p=12$ leads to sparse activations without zeroing the whole activations.}

\noindent{\bf Training The Sparse Node Activation Loss\quad} Fig.\ref{fig:ber_snr}(d) provides the BER curve during training with the sparse node activation only (i.e. without the cross-entropy loss and the knowledge distillation loss). We use the BCH (63,45) code with a right-regular parity check matrix, and we plot the BER curve after each epoch. As we can observe, after two epochs the training improves the BER curve drastically at a high SNR regime. However, for a low SNR regime, it gives high degradation in the BER performance. We can also observe that after three epochs, the training explodes and the BER curve has a worse performance compared to the previous epochs. Moreover, after four epochs, the BER curve has an inferior performance compare to all previous epochs. 
The phenomena of training exploding is reasonable since the neural decoder is trained to output zero activation at the variable and check nodes which gives the wrong estimation to the transmitted codeword. However, we can observe that during training the sparse node activation loss leads to high improvement of BER in a high SNR regime. 
\vspace{-0.25cm}
\section{Conclusion}
We propose new learning loss terms for neural belief propagation decoding. The new learning loss terms use sparse constraints on the variable and check nodes, furthermore, it uses a teacher network to guide the training of the neural network. The proposed method shows improvement of up to $1.1dB$ in the high SNR regime. Moreover, the method has the same run-time complexity and the same model size as the baseline methods. The sparse constraint on neural decoders is an interesting research direction and it is left for future work to discover whether other sparse constraints (for example - weights) can improve the performance of the neural decoder. The usage of Teacher-Student learning for Neural Belief Propagation can also be beneficial for other applications, such as online learning (when the ground truth is unknown). For example, it would be interesting to see the improvement when the student network is trained with changing real-world data. Furthermore, it also left for future research to apply the proposed method to the neural Polar factor graph decoder \cite{xu2017improved}.

\section*{Acknowledgments}
We thank Uriel Singer and Monica Broido for the helpful discussion. The contribution of Eliya Nachmani is part of a Ph.D. thesis research conducted at Tel Aviv University.

\bibliographystyle{IEEEtran}
\bibliography{graphs}

\vfill

\end{document}